\documentclass[runningheads]{llncs}
\usepackage{graphicx}

\AtBeginDocument{%
  \providecommand\BibTeX{{%
    \normalfont B\kern-0.5em{\scshape i\kern-0.25em b}\kern-0.8em\TeX}}}
\usepackage{algorithm2e}

\begin{document}

\title {Porting of eChronos RTOS on RISC-V Architecture}


\author{Shubhendra Pal Singhal\inst{1} \and
M. Sridevi\inst{1} \and
N Sathya Narayanan\inst{2} \and M J Shankar Raman \inst{2}}
\authorrunning{Shubhendra Pal Singhal}
%
\institute{Department of Computer Science and Engineering\\National Institute of Technology, Tiruchirappalli\\
\email{106116088@nitt.edu }\\ 
\email{msridevi@nitt.edu}\and
Department of Computer Science and Engineering\\Indian Institute of Technology, Madras\\
\email{sathya281@gmail.com}\\
\email{mjsraman@gmail.com }}
\maketitle            

\begin{abstract}
eChronos is a formally verified Real Time Operating System (RTOS) designed for embedded micro-controllers. eChronos was targeted for tightly constrained devices without memory management units. Currently, eChronos is available on proprietary designs like ARM, PowerPC and Intel architectures. eChronos is adopted in safety critical systems like aircraft control system and medical implant devices. eChronos is one of the very few system software's not been ported to RISC-V. RISC-V is an open-source Instruction Set Architecture (ISA) that enables new era of processor development. Many standard Operating Systems, software tool chain have migrated to the RISC-V architecture. According to the latest trends\cite{eChronos14}, RISC-V is replacing many proprietary chips. As a secure RTOS, it is attractive to port on an open-source ISA. SHAKTI and PicoRV32 are some of the proven open-source RISC-V designs available. Now having a secure RTOS on an open-source hardware design, designed based on an open-source ISA makes it more interesting. In addition to this, the current architectures supported by eChronos are all proprietary designs \cite{eChronos2}, and porting eChronos to the RISC-V architecture increases the secure system development as a whole. This paper, presents an idea of porting eChronos on a chip which is open-source and effective, thus reducing the cost of embedded systems. Designing a open-source system that is completely open-source reduces the overall cost, increased the security and can be critically reviewed. This paper explores the design and architecture aspect involved in porting eChronos to RISC-V. The authors have successfully ported eChronos to RISC-V architecture and verified it on spike\cite{eChronos13}. The port of RISC-V to eChronos is made available open-source by authors\cite{eChronos8}. Along with that, the safe removal of architectural dependencies and subsequent changes in eChronos are also analyzed.

\end{abstract}

\section{Introduction}
Porting is the process of converting the software for the another architecture. A software has many exceptions and forms of programming specific to the architecture like memory or register allocation. To be able to run such executable file, we need porting. There are many OS porting techniques available in the literature \cite{eChronos4} that follow trial and error approach. In this paper, we have used the hit and trial approach for porting eChronos on an open-source architecture RISC-V\cite{eChronos11,eChronos1}.\newline

\subsection{eChronos}
The eChronos RTOS is a real-time operating system (RTOS)  \cite{eChronos2}. It is intended for tightly resource-constrained devices without memory management units and virtual memory support. Also, RTOS code base is designed to be highly modular so that only the minimal amount of code necessary is ever compiled into a given system image\cite{eChronos2}.
 The eChronos RTOS is highly configurable and functional as shown in FIG 1\cite{eChronos}.

\begin{figure}
\begin{center}
\includegraphics[scale = 0.5]{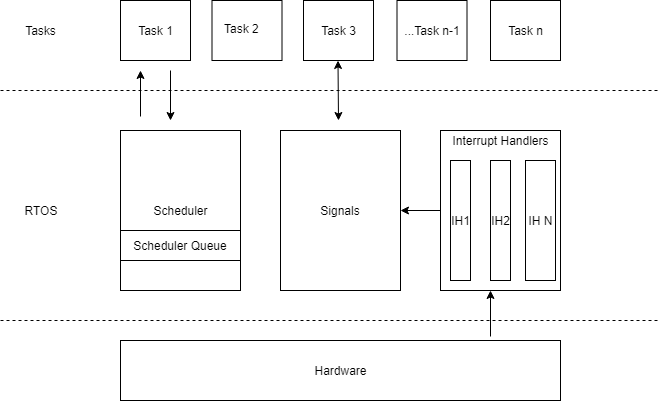}
\end{center}{}
\caption{eChronos RTOS functionality}
\end{figure}{}

\subsection{RISC-V}
RISC-V is an open-source hardware ISA based on Reduced Instruction Set Computer (RISC) principles. RISC-V is a layered and extensible open-source ISA which means that it can support the implementation of well defined extensions for a given application. RISC-V offers a very flexible usage of instruction set which adds to upto RISC-V becoming more popular\cite{eChronos14}. RISC-V is modular and flexible, which reduces the effort to develop the ancillary software for a specific processor. 
\subsection{Porting an operating system}
Porting of OS is a time consuming process as it involves a changes to big and complex programs. Porting every specific program seems to be a bit irrational. So the solution to such a complex problem are the modern day compilers, which translates the high level language program to a platform independent code unlike the traditional compilers translating the code directly to the machine code. The intermediate language can then execute all programs and it gets translated into a sequence of machine code by a code generator to create a executable file. The use of intermediate code enhances the portability of the compiler, because only the machine dependent code (the interpreter or the code generator) of the compiler needs to be ported instead of porting the program itself. The remaining part of the code in the compiler can be treated as an intermediate code and then processed by the ported code interpreter. This reduces design efforts, because the machine independent code just needs to be developed only once to create a portable intermediate code. An interpreter is less complex to code and it follows a certain algorithm avoiding the need to be specifically port for every single program \cite{eChronos3}. A software can be compiled and linked from source code for various operating systems and architectures. Real Time Operating System files are all architecture dependent. A small mistake in port files will lead to an easy system crash. \cite{eChronos4,eChronos7}. \newline

\section{Porting eChronos}

The subsequent sections explain the step-wise process involved in the porting of eChronos for RISC-V architecture.

\subsection{Prerequisites for porting eChronos on RISC-V }

The following are the prerequisites

\begin{itemize}
\item[1] A Linux operating system running on personal computer with 4GB RAM.
\item[2] A GNU based environment, that has all necessary software packages mentioned in \cite{eChronos13}.

\end{itemize}

\subsection{Platform set up for eChronos}

The following steps need to be done to run eChronos on an emulator,

\subsubsection{Installation of RISC-V emulator}
Follow the following steps that are listed below for the installation of RISC-V tools \cite{eChronos6,eChronos12,eChronos10}: \newline \newline
\noindent
\begin{itemize}
    
\item[1] \textbf{The following packages are required for the RISC-V tools:}\newline
\noindent
autoconf automake autotools-dev curl libmpc-dev libmpfr-dev libgmp-dev libusb-1.0-0-dev gawk build-essential bison flex texinfo gperf libtool patchutils bc zlib1g-dev device-tree-compiler pkg-config \newline
libglib2.0-dev zlib1g-dev libpixman-1-dev \newline
\noindent
   \item[2] \textbf{ Set up a directory for the RISC-V tools, and the RISC-V environment variable, which we refer to in future steps.}\newline
\noindent
     mkdir RISC-V\\
     cd RISC-V\\
     export RISC-V=$\{PWD\}$ \newline
     
\noindent
  \item[3] \textbf{  Get the RISC-V tools}\newline
    \noindent
cd {$RISC-V$}\\
  git clone https://github.com/RISC-V/RISC-V-git\\
cd RISC-V-tools\\
     git submodule update --init --recursive\\
     
     \noindent
   \item[4] \textbf{ Get RISC-V qemu sources, and build them:}\newline
\noindent
     cd ${RISC-V}/RISC-V-tools\\
     $git clone https://github.com/heshamelmatary/RISC-V-qemu.git$\\
     $cd RISC-V-qemu$\\
     $git checkout sfence$\\
     $git submodule update --init dtc$\\
     $./configure --target-list=RISC-V64-softmmu,RISC-V32-softmmu --prefix=${RISC-V}$\\
     make -j8 \&\& make install\\
\noindent \newline
 \item[5] \textbf{   Build the 64-bit toolchain}\newline
\noindent
     cd {RISC-V}/RISC-V-tools\\
     sed -i 's/build\_project RISC-V-gnu-toolchain --prefix=\$RISC-V/build\_project RISC-V-gnu-toolchain --prefix=RISC-V --with-arch=rv64imafdc --with-abi=lp64'\\
     ./build.sh\\
\noindent \newline
 \item[6] \textbf{Installation of eChronos}\newline
\noindent
Download and install the eChronos\cite{eChronos14}.
\end{itemize}
\section{eChronos system design}

\begin{figure}
\begin{center}
\includegraphics[scale = 0.6]{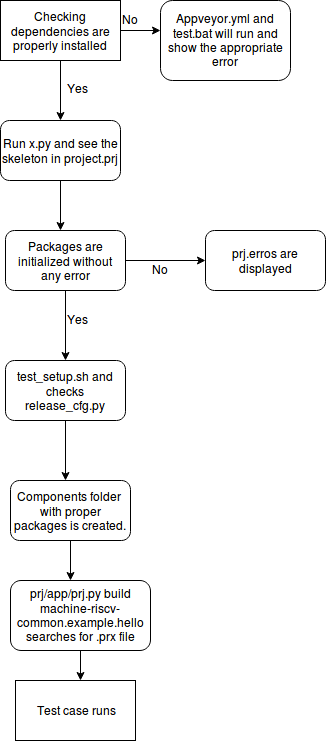}
\end{center}
\caption{Flow Diagram of eChronos}
\end{figure}

The FIG.2 shows the system diagram of eChronos. The following steps explain FIG 2\cite{eChronos9}. that help eChronos produce an executable file (of the test program)\cite{eChronos2}:\newline

\begin{itemize}
\item[1] Ensure that there is no break in the packages installed earlier and the dependencies are installed properly. A yml file is created for the purpose of testing the version of the dependencies installed and whether they are suitable for running eChronos or not? : e.g. running 3.4 version of python needs to be verified.
\item[2] $/$x.py build packages executes x.py file which uses the skeleton of the system as a reference for the order of installation. The following output files will be produced by x.py :
\begin{itemize}
\item release$/prj-<version>$.zip 
\item release$/<rtos-foo>-<version>$.zip
\item release$/<build-name>-<version>$.zip 
\end{itemize}

\item[3] If the installation shows any error related to traceback like Traceback (most recent call last):
\vspace{.5cm}

 File "prj/app/prj.py", line 1293, in start
    sys.exit(main()), then the installation or dependencies are not installed properly. 
    Follow the earlier steps again properly.
\vspace{.5cm}

\item[4] After prj errors are removed, we have to ensure that the class standard cfg.py should contain the RISC-V architecture. Simultaneously, spike\cite{eChronos13} emulator is installed. The test\_setup.sh script downloads the package eChronos from Github. It contains travis to ensure that if package is missing then the alternative can be downloaded i.e. the posix version\cite{eChronos2}.\newline
\item[5] Components folder:\\
api-conditions,  context-switch-RISC-V,  docs,  error,  interrupt-event-RISC-V,  message-queue, rigel, rigel.py,  stack-RISC-V,  task,  timer-RISC-V are the necessary components that will be created once the script runs properly.\newline
\item[6] After ensuring a proper installation of eChronos, we test our sample program. prj/app/prj.py build $<example\_name>$ command runs the test case on eChronos RTOS and produces the binary file which can be tested on spike.\newline
\item[7] The block "Test case runs" is explained in detail in section 4.1.3.

\end{itemize}

\subsection{Challenges encountered in porting}
\noindent Porting requires the understanding of the flow of data and the subsequent changes in the system related to architecture. As eChronos is architecture dependent, there were many changes pertaining to specificity of the architecture like : General architecture similar to 8086 consists of extra data segment by default, but in case of RISC-V architecture, it has to be declared explicitly. The default edata (extra data segment) initialization is missing from RISC-V architecture so the edata declaration had to be removed from every file in eChronos.\cite{eChronos5}. Likewise, there were some architecture dependent terms which had to be modified, so as to provide support to eChronos for RISC-V.

\section{Step-wise porting of eChronos on RISC-V}
\noindent eChronos initially could run only one program(test case) i.e. concurrency of task A and task B. eChronos could not run any generic program like "Hello World". Porting eChronos on RISC-V, thus became difficult because RISC-V cannot support the execution of concurrent programs. So porting eChronos on RISC-V meant that we have to change the test case to a generic one which RISC-V can support. Thus, eChronos was first changed to support generic programs like "Hello World" and then the subsequent execution of its exe file on the RISC-V could intimate whether eChronos is ported successfully or not.    
\noindent

\subsection{Changing test case in eChronos}

\subsubsection{Hit and trial method of porting}

\noindent The components section of eChronos consists of libraries like mutex, context switching and other concurrency supportive packages which are not used by our sample program. We removed all the unnecessary packages and removed their declarations from all the files like x.py core configuration, prj.py file beacuse "build packages" command would take unnecessary time executing irrelevant packages otherwise.\newline
After ensuring correct set of packages like rigel, api-conditions, errors, doc etc. we try to modify the test case. Test case has only one external dependency i.e. print statement. "printf" could used as one of the options but then gcc support for eChronos needs to be ensured which leads to multiple dependency issues and these issues might be difficult to track. Also, we need to port eChronos with minimal set of library dependencies and thus, we configured the print function\cite{eChronos8} by using write() command of stdio library which is comparatively a less complex function than printf function of gcc. Thus, a test program with least dependency is programmed and now, we have to produce its executable file to ensure that every interlinked package is working properly. This implies that if its executable file is generated, then removal of unnecessary packages was a success  plus the test case is running properly on eChronos else we have to again solve the package error and keep on trying(hit and trial). Further, if risc-v linker along with this executable file, generates a system dump (which executes properly on spike too) then eChronos is successfully ported on RISC-V.   

\subsubsection{Initialization of packages}
\noindent The main principle of porting is to understand what lines of code to modify in accordance with the architecture and what dependencies needs to be taken into consideration. In this case the prj tool written in python is the initial script which initializes all the packages present and creates the required directories. When prj tools runs in eChronos, project.prj is the skeleton file which runs x.py file. It states all the dependencies of packages that is required by the system to function properly. There are many components like errors, api-conditions etc which are addressed by eChronos and needs to be initialized. x.py helps in initialization of these packages. \\
.$/$x.py build packages\\
The command .$/$x.py will initialize the packages that are mentioned in skeleton of your system in x.py file. For instance : For RISC-V, rigel is the system name which has packages like error, test case etc. mentioned in x.py file, which helps in initial creation of sub-directories. This helps to create the build structure in a modular fashion where every module consists of .exe files of different packages initialized in x.py file.

\subsubsection{Running an application on eChronos}

\noindent prj/app/prj.py build machine-RISC-V-common.example.hello is the command that runs the test case by first creating an exe file of the test case.
After creating .exe files of packages using build command, we run test case using these packages. 
This exe file created by OS is architecture dependent and may raise some errors if OS files are corrupted or not ported correctly.\\

\begin{figure}
\begin{center}
\includegraphics[scale = 0.7]{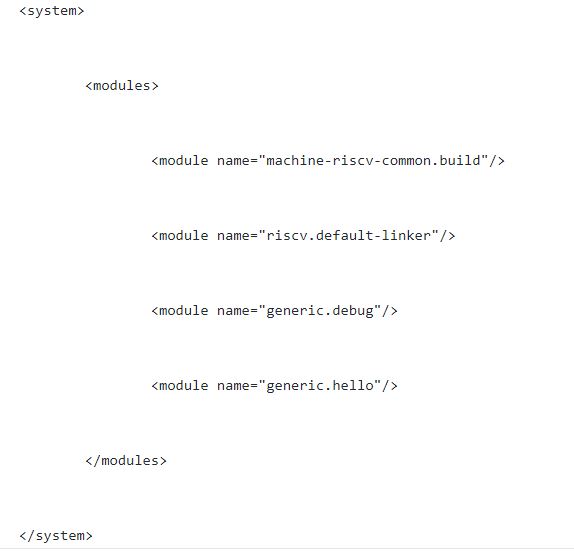}
\end{center}{}
\caption{machine-riscv-common.example.hello.prx}
\end{figure}{}

Test case is stored in the hello.c file. machine-RISC-V-common.example.hello.prx consists of skeleton of how the test case should run as shown in Fig3. It consists of the modules names which needs to be executed in the sequential order. 
This command searches for .prx file which contains all the files to be used for running your program stored in the system. \newline \newline We will illustrate this using our model rigel.\newline 
1.) .prx file instructs the system to compile debug.c in generic folder, hello.c in generic folder, default-linker.py in RISC-V folder and build.py as mentioned in machine-RISC-V-common.example.hello.prx file.\\ 
2.) RISC-V.default-linker means that we should go to the RISC-V folder and run default-linker.py file. One thing must remembered that only .c, .py and other high level language files can run in the .prx file.

\subsubsection{Data flow of a running application}

\noindent This section tries to explain the data flow in eChronos, when a test case run. The file "debug.h" consists of initialization of function : debug\_print and debug.c consists of its declaration. The hello.c i.e. the test case uses debug.h as its header file and uses the function debug\_print to print "Hello world". This requires the compilation of header files. \newline
1.) build.py file takes the system files and some other helper files like debug.h to compile, assemble and run your hello.c file in generic folder.\newline
2.) hello.c file just wants to print "Hello World". So for printing the command printf is not used, rather we have created the debug.c file where the print command has been declared. printf command is not used because it is a part of gcc library which is yet not ported to RISC-V. So a generic print statement is declared in debug file.\newline
3.) The compilation of all libraries is over. Rigel model along with its components is build and the libraries used in the test case are also compiled. Only proper linking of all these files is left. The linker file default.ld is created where the linking occurs according to RISC-V architecture. Linker file of different architectures was referred while making default.ld. This linker file assumed "edata" one of the variables to be declared by default but RISC-V does not support any default declarations. So we changed "edata=null" and linking errors were solved. Thus, all the compiled files were linked. 

\subsection{Running test case on spike(RISC-V emulator)} 
\noindent As soon as you hit the first command, the out folder is created with rtos-$<systemname>$ folder. Continuing further, if the second command runs with no error then the system image will be created in the $<program-name>$ folder. This, then needs to be executed on your respective architecture as in our case it is on spike, RISC-V emulator\cite{eChronos3,eChronos13}. 
\section{Results}
\noindent Running the system dump of hello.c file gave an output : "Hello World" assuming the starting address of execution of program was 10000 on the architecture. RISC-V emulator\cite{eChronos3} has been designed in such a way that the address for the program needs to be specified above 10000 only. But its just a design issue which will not be of any problem in the real chip SHAKTI\cite{eChronos11}. This indicates that the executable file produced by eChronos successfully ran on spike(RISC-V emulator). The output of sample program on eChronos is available (open source)\cite{eChronos8}. 

Additions of any critical programs will be just a further extension of ported Hello World program provided the RISC-V supports that functionality of the program. Supposedly, a program which needs to determine the shortest path needs to be ported on RISC-V using eChronos. Program would consist of the same logic but instead of using gcc functions we will be using generic functions as seen with print in debug.h. This ensures a safe porting technique with no dependency on outer(alien) packages. Thus, eChronos is ported for RISC-V, for the base program hello world. Any additions to eChronos will be an extended version of the ported hello.c file with some extra headers files(declaring some predefined functions).

\section{Conclusion}
\noindent eChronos real-time operating system is ported to the RISC-V architecture and successfully executed on spike. The sample program hello.c has only one external dependency i.e. print. This is the base level porting for RISC-V that can be used as a reference for further adaptability of complex programs. Extensive porting of libraries in eChronos can be done by modifying the files in the same way as the sample program file\cite{eChronos15}. 
\bibliographystyle{splncs04}
\bibliography{sample-base}

\end{document}